\documentclass[oldversion]{aa} 
\usepackage{graphicx}
\usepackage{txfonts}

\def\etal{{et al. \rm}}

\def\bcma{\object{$\beta$ CMa}}
\def\xcma{\object{$\xi^1$ CMa}}
\def\neri{\object{$\nu$ Eri}}
\def\dcet{\object{$\delta$ Cet}}
\def\voph{\object{V2052 Oph}}
\def\vcen{\object{V836 Cen}}
\def\twelvelac{\object{12 Lac}}
\def\gpeg{\object{$\gamma$ Peg}}
\def\bcep{\object{$\beta$ Cep}}
\def\gori{\object{$\gamma$ Ori}}
\def\ecma{\object{$\epsilon$ CMa}}
\def\piori{\object{$\pi^4$ Ori}}
\def\15cma{\object{15 CMa}}
\def\icma{\object{$\iota$ CMa}}
\def\bcru{\object{$\beta$ Cru}}
\def\toph{\object{$\theta$ Oph A}}
\def\zcas{\object{$\zeta$ Cas}}
\def\tsco{\object{$\tau$ Sco}}
\def\tsco{\object{$\tau$ Sco}}

\begin{document}
   \title{
Nitrogen enrichment, boron depletion and magnetic fields in slowly-rotating B-type dwarfs
}
   \titlerunning{Nitrogen enrichment, boron depletion and magnetic fields in slowly-rotating B-type dwarfs}

   \author{T. Morel
         \inst{1}
          \fnmsep\thanks{Present address: Institut d'Astrophysique et de G\'eophysique, Universit\'e de Li\`ege, All\'ee du 6 Ao\^ut, B\^at. B5c, 4000 Li\`ege, Belgium}
          \and
          S. Hubrig
          \inst{2}
          \and
          M. Briquet
          \inst{1}
          }

   \offprints{Thierry Morel, \email{morel@astro.ulg.ac.be}.}

   \institute{Katholieke Universiteit Leuven, Departement Natuurkunde en Sterrenkunde, Instituut voor Sterrenkunde, Celestijnenlaan 200D, B-3001 Leuven, Belgium
        \and
        European Southern Observatory, Casilla 19001, Santiago, Chile}

   \date{Received 6 November 2007 / Accepted 15 January 2008}

\abstract{Evolutionary models for massive stars, accounting for rotational mixing effects, do not predict any core-processed material at the surface of B dwarfs with low rotational velocities. Contrary to theoretical expectations, we present a detailed and fully-homogeneous, NLTE abundance analysis of 20 early B-type dwarfs and (sub)giants that reveals the existence of a population of nitrogen-rich and boron-depleted, yet intrinsically slowly-rotating objects. The low-rotation rate of several of these stars is firmly established, either from the occurrence of phase-locked UV wind line-profile variations, which can be ascribed to rotational modulation, or from theoretical modelling in the pulsating variables. The observational data presently available suggest a higher incidence of chemical peculiarities in stars with a (weak) detected magnetic field. This opens the possibility that magnetic phenomena are important in altering the photospheric abundances of early B dwarfs, even for surface field strengths at the one hundred Gauss level. However, further spectropolarimetric observations are needed to assess the validity of this hypothesis.} 


   \maketitle
%

\section{Introduction} \label{sect_intro}
Fast rotation is one of the most distinctive features of massive stars and may give rise to large-scale fluid motions in the stellar interior which dredge up a significant amount of CNO-processed material from the stellar core to the surface (Heger \& Langer \cite{heger_langer}; Meynet \& Maeder \cite{meynet_maeder00}). On the other hand, such mixing processes can transport the boron surviving close to the stellar surface, after the star has evolved off the zero age main sequence (ZAMS), to deeper layers where this fragile element will quickly be destroyed. Therefore, the B and CNO surface abundances of OB stars are powerful probes of rotation-related mixing phenomena throughout the entire stellar interior and at different evolutionary phases. As such, they can be used to constrain theoretical models including rotational effects, such as meridional currents or shear instabilities, and to ultimately calibrate some poorly-known quantities, such as the diffusion coefficients for the transport of the chemical elements. Of particular relevance for the present work is the recent recognition that dynamo-generated magnetic fields could also significantly alter the photospheric CNO content (Maeder \& Meynet \cite{maeder_meynet}; Heger \etal \cite{heger}). 

The evolutionary models mentioned above do not predict any detectable nitrogen overabundance (i.e. above 0.1--0.2 dex) in B dwarfs with rotation rates below $\sim$100 km s$^{-1}$. However, it has now been known for a long time (but often overlooked) that some apparently slowly-rotating objects actually present relatively large quantities of N-enriched material at their surface, with overabundances reaching up to a factor three (e.g. Kilian \cite{kilian92}; Gies \& Lambert \cite{gies_lambert}). Such an extra amount of deep mixing in OB dwarfs would clearly have consequences on their post main-sequence evolution and other related, important issues (e.g. stellar populations in external galaxies). 

We have recently revealed the same peculiarity in a number of $\beta$ Cephei stars, which have been inferred from asteroseismic or line-profile variation studies to be {\em intrinsically} very slow rotators. As an illustration, the B1.5--B2 subgiant \dcet \ has a true rotation rate of at most 28 km s$^{-1}$, as deduced from the detection of a rotationally-split multiplet in {\em MOST} data (Aerts \cite{aerts06}), yet exhibits an unexpected N excess reaching a factor about four (Morel \etal \cite{morel}; hereafter M06). This is puzzling when considering, for instance, that no contamination from CN-processed material has been reported in the literature for some fast-rotating analogs in low-metallicity environments where, in addition, the abundance changes should be more easily detectable (e.g. several B1.5 subgiants in the LMC with equatorial rotational velocities exceeding $\sim$130 km s$^{-1}$; Korn \etal \cite{korn}). To shed more light on this issue, here we extend the M06 study and present an NLTE abundance analysis of a sample of 20 dwarfs/(sub)giants comprising stars not established as pulsating variables, with special emphasis on the possible role played by magnetic fields to shape the CNO surface abundance patterns of early B-type stars during the early phases of their evolution. 

\section{Target selection and observational material} \label{sect_obs}
We have complemented the sample analyzed by M06 with six early B-type dwarfs defined as boron-poor from the analysis of the \ion{B}{iii} $\lambda$2066 UV resonance line (Brooks \etal \cite{brooks}; Mendel \etal \cite{mendel}; Proffitt \& Quigley \cite{proffitt_quigley}; Venn \etal \cite{venn02}). All stars have boron abundances at least one order of magnitude lower than the meteoritic value. The studied sample includes {\em all} boron-depleted B stars known to date. For comparison purposes, we also analyzed the boron-normal star \bcru \ (Proffitt \& Quigley \cite{proffitt_quigley}). 

In the case of \icma \ and \ecma \ we made use of a mean FEROS spectrum created by averaging 6 and 4 individual exposures closely taken in time, respectively. These data were obtained in April 2005 with the ESO/MPI 2.2m Telescope and were kindly provided by P. Dufton. Echelle spectra of \gori \ (mean of two exposures) and \object{HD 36591} were retrieved from the ELODIE archives\footnote{See {\tt http://atlas.obs-hp.fr/elodie/}.}. Snapshot spectra of \piori \ and \15cma \ (4 and 3 exposures obtained over a timespan of about one week, respectively) were obtained in March 2007 with the CORALIE spectrograph attached to the ESO/Euler 1.2m Telescope at La Silla. For the $\beta$ Cephei star \bcru, we used a mean CORALIE spectrum created by co-adding a set of 49 individual exposures obtained over the period 21--29 March 2007. 

In addition, we discuss here previous results obtained following exactly the same procedures for two slowly pulsating B stars (\zcas \ and \object{HD 85953}; Briquet \& Morel \cite{briquet_morel}), the magnetic star \tsco \ (Hubrig et al., in prep.) and the $\beta$ Cephei star \toph \ (Briquet \etal \cite{briquet}). It is important to emphasize that the performed abundance analysis is identical in all respects for all the stars discussed in the present paper. The abundance data presented here thus constitute a fully homogeneous database suitable for a comparative study. This point is essential considering the mild nitrogen excesses detected, as will be shown below.

The majority of the stars in our sample have been the subject of sensitive spectropolarimetric observations, which led to the detection of a weak-longitudinal, magnetic field in \gpeg \ (Butkovskaya \& Plachinda \cite{butkovskaya}), \zcas \ (Neiner \etal \cite{neiner03a}), \xcma \ (Hubrig \etal \cite{hubrig06}), \object{HD 85953} (Hubrig \etal \cite{hubrig06}), \tsco \ (Donati \etal \cite{donati06b}), \voph \ (Neiner \etal \cite{neiner03b}) and \bcep \ (Henrichs \etal \cite{henrichs}). Very recent observations with FORS1/VLT not only confirm the field detection in \xcma, but also reveal a weaker field at the one hundred Gauss level in \dcet, \15cma \ and \ecma \ (Hubrig et al., in prep.). On the other hand, null detections have been reported from observations with similar sensitivity limits for \neri, \bcma, \bcru, \vcen \ and \toph \ (Schnerr \etal \cite{schnerr06a}; Hubrig \etal \cite{hubrig06}; Hubrig et al., in prep.). The low-sensitivity circular polarization observations of \twelvelac \ by Rudy \& Kemp (\cite{rudy_kemp}) rule out the existence of a magnetic field at the kG level. Table~\ref{tab_magnetic_fields} summarizes the basic magnetic properties of the program stars and provides, for those with detections, the rms longitudinal field, $\overline{\left< B_l \right>}$, and the reduced $\chi^2$ of the measurements. The former is a measure of the intrinsic strength of the field, while the latter can be used to assess the reliability of the detection (see, e.g., Hubrig \etal \cite{hubrig06}). These quantities are defined from $N$ individual measurements of a mean longitudinal magnetic field, $\left<B_l\right>_i$, with an associated uncertainty $\sigma_i$, as:

\begin{equation}
\overline{\left< B_l \right>} = \left( \frac{1}{N} \sum^{N}_{i=1} \left< B_l \right> ^2_i \right)^{1/2}
\end{equation}

and

\begin{equation}
\chi^2/N = \frac{1}{N} \sum_{i=1}^N \left( \frac{\left< B_l \right>_i}{\sigma_i} \right)^2
\end{equation}

It should be noted that the amount of spectropolarimetric data collected --- and therefore the robustness of the magnetic field (non)detection --- varies dramatically from one star to another: while few targets have been the subject of intensive, multi-epoch observing campaigns, others only have a handful of snapshot observations available. This important point will be discussed further in Sect.~\ref{sect_fields}.

\begin{table}
\centering
\caption{Longitudinal magnetic field properties of the program stars. Note that the rms longitudinal field values, $\overline{\left< B_l \right>}$, are only indicative in case of very few measurements (e.g. \15cma \ and \ecma). A field is considered undetected if it does not have a significance of at least 3$\sigma$.}
\label{tab_magnetic_fields}
\begin{tabular}{rccccc} \hline\hline
\multicolumn{1}{l}{HD number} & Usual name & $\overline{\left<B_l\right>}$ [G] & $N$ & $\chi^2/N$ & Ref.\\\hline
     886 & \gpeg       & 20 & 17 & 3.8  & 1\\
    3360 & \zcas       & 38 & 118 & 2.0 & 2\\ 
   16582 & \dcet       & 85 &   9 & 9.5 & 3 \\
   29248 & \neri       & \multicolumn{3}{c}{Not detected} & 3,4,5\\
   30836 & \piori      & \multicolumn{3}{c}{Not observed}\\
   35468 & \gori       & \multicolumn{3}{c}{Not observed}\\
   36591 &             & \multicolumn{3}{c}{Not observed}\\
   44743 & \bcma       & \multicolumn{3}{c}{Not detected} & 3,4\\
   46328 & \xcma       & 287 &  11 & 478 & 3,4\\
   50707 & \15cma      & 144 &    2 &  19 & 3  \\
   51309 & \icma       & \multicolumn{3}{c}{Not observed}\\
   52089 & \ecma       & 149 &    2 &  14 & 3 \\
   85953 &             & 102 &    5 &  6.9 & 3,4\\
111\,123 & \bcru       & \multicolumn{3}{c}{Not detected} & 4\\
129\,929 & \vcen       & \multicolumn{3}{c}{Not detected} & 3,4\\
149\,438 & \tsco       & 42 & 35 & 93 & 6\\
157\,056 & \toph       & \multicolumn{3}{c}{Not detected} & 4\\
163\,472 & \voph       & 262 & 112 & 1.3 & 7\\
205\,021 & \bcep       & 80 & 124 & 8.4 & 8\\
214\,993 & \twelvelac  & \multicolumn{3}{c}{Not detected} & 9\\
\hline
\end{tabular}
\begin{flushleft}
Key to references: [1] Butkovskaya \& Plachinda (\cite{butkovskaya}); [2] Neiner \etal (\cite{neiner03a}); [3] Hubrig et al., in prep.; [4] Hubrig \etal (\cite{hubrig06}); [5] Schnerr \etal (\cite{schnerr06a}); [6] Donati \etal (\cite{donati06b}); [7] Neiner \etal (\cite{neiner03b}); [8] Henrichs \etal (\cite{henrichs}); [9] Rudy \& Kemp (\cite{rudy_kemp}).\\
\end{flushleft}
\end{table}

\section{Methods of analysis and sample characteristics} \label{sect_analysis}
The non-local thermodynamic equilibrium (NLTE) abundances of He, C, N, O, Mg, Al, Si, S and Fe were calculated using the latest versions of the line-formation codes DETAIL/SURFACE and plane-parallel, fully line-blanketed Kurucz atmospheric models (Kurucz \cite{kurucz93}). A standard, iterative scheme is first used to derive the atmospheric parameters purely on spectroscopic grounds: $T_{\rm eff}$ is determined from the \ion{Si}{ii/iii/iv} ionization balance, $\log g$ from fitting the collisionally-broadened wings of the Balmer lines, and the microturbulent velocity, $\xi$, from requiring the abundances yielded by the \ion{O}{ii} features to be independent of the line strength. The ionic species \ion{C}{iii}, \ion{N}{iii} and \ion{S}{iii} are poorly treated in our model atoms and only a few spectral features are measurable, but there is no evidence in the B0--B1 stars for systematic discrepancies with the abundances yielded by the much more numerous lines of lower ionization stages (namely \ion{C}{ii}, \ion{N}{ii} and \ion{S}{ii}). All stars are slow rotators (see below) and classical curve-of-growth techniques were used to determine the abundances using the equivalent widths (EWs) of a set of unblended lines. The abundance uncertainties take into account both the line-to-line scatter and the errors arising from the uncertainties in the atmospheric parameters. Finally, we uniformly derived the projected rotational velocities, $\Omega R \sin i$, by comparing the profiles of a set of isolated \ion{O}{ii} lines with a grid of rotationally-broadened synthetic spectra. The reader is referred to M06 for complete details. 

To correct for the contamination of the spectrum by the secondary in the single line binary \bcru, we based our analysis on synthetic, composite spectra assuming the following parameters for the companion: $T_{\rm eff}$=22\,000 K, $\log g$=4.0 dex [cgs] (Aerts \etal \cite{aerts98}) and a microturbulent velocity, $\xi$=5 km s$^{-1}$, typical of B-type dwarfs. We considered that the flux ratio between the two components is 8\% in the optical band (Hanbury Brown \etal \cite{hanbury_brown}; Popper \cite{popper}) and that the secondary has a chemical composition typical of OB dwarfs in the solar vicinity (Daflon \& Cunha \cite{daflon_cunha}; the value for iron was taken from M06). To examine the sensitivity of our results to these various assumptions, we have repeated the abundance analysis after varying the adopted effective temperature, surface gravity, chemical composition, and luminosity of the secondary within the range of plausible values. Namely, we assumed in turn: $T_{\rm eff}$=20\,000 K, $\log g$=3.7 dex [cgs], the abundances of all the metals enhanced by 0.3 dex relative to solar and a flux ratio in the optical between the two components amounting to up to 12\%, while keeping the other parameters unchanged. As can be seen in Table~\ref{tab_sensitivity}, taking the companion into account (regardless of its characteristics) has a negligible impact on the abundance results. The same conclusion holds in the case of the  single line binary \toph \ (see Briquet \etal \cite{briquet}). The stars \gpeg \ and \bcep \ are associated with faint secondaries (Roberts \etal \cite{roberts}; Schnerr \etal \cite{schnerr06b}), and their contribution was neglected. The giant \piori \ is also part of a massive binary (as confirmed by the radial velocity changes observed in our own dataset) with an orbital period of about 9.5 d (Pourbaix \etal \cite{pourbaix}), but the lack of information about the secondary forced us to treat it as a single star (Gies \& Lambert \cite{gies_lambert} classify \piori \ as B2 III + B2 IV, but no details are given). 

\begin{table*}
\centering
\caption{Sensitivity of the derived abundances of $\beta$ Cru on the assumed properties of the secondary. We quote the abundance differences with the values provided in Tables \ref{tab_abundances_cno} and \ref{tab_abundances}.}
\label{tab_sensitivity}
\begin{tabular}{lcccccc} \hline\hline
   & Ionic species used    & Without companion & \multicolumn{4}{c}{With companion}\\
        &                  & & $\Delta$$T_{\rm eff}$=--2000 K & $\Delta$$\log g$=--0.3 dex & $\Delta\log \epsilon$=+0.3 dex & flux ratio=12\% \\
\hline
He/H                      & \ion{He}{i}                               &   +0.005 &  +0.000 &   +0.002 &   +0.000 & --0.003\\ 
$\Delta\log \epsilon$(C)  & \ion{C}{ii} (+\ion{C}{iii} $\lambda$4187) &    +0.03 &   +0.02 &   --0.01 &   --0.05 &  --0.02\\ 
$\Delta\log \epsilon$(N)  & \ion{N}{ii} (+\ion{N}{iii} $\lambda$4634) &    +0.00 &   +0.02 &   --0.01 &   --0.02 &   +0.00\\ 
$\Delta\log \epsilon$(O)  & \ion{O}{ii}                               &  -- 0.04 &   +0.01 &   --0.01 &   --0.01 &   +0.02\\ 
$\Delta\log \epsilon$(Mg) & \ion{Mg}{ii}                              &    +0.03 &  --0.02 &    +0.00 &   --0.02 &  --0.02\\ 
$\Delta\log \epsilon$(Al) & \ion{Al}{iii}                             &    +0.01 &   +0.02 &    +0.00 &   --0.03 &   +0.00\\ 
$\Delta\log \epsilon$(Si) & \ion{Si}{iii/iv}                          &   --0.04 &   +0.01 &    +0.00 &   --0.01 &   +0.02\\ 
$\Delta\log \epsilon$(S)  & \ion{S}{iii}                              &   --0.03 &   +0.01 &   --0.01 &   --0.01 &   +0.01\\ 
$\Delta\log \epsilon$(Fe) & \ion{Fe}{iii}                             &    +0.00 &   +0.01 &   --0.01 &   --0.03 &   +0.00\\ 
\hline
\end{tabular}
\end{table*}

The adopted atmospheric parameters are provided in Table~\ref{tab_parameters}. The same parameters, within the errors, are also obtained for all studied stars when analyzing our data with the unified code FASTWIND (Puls et al. \cite{puls}) and (semi-)automatic line-profile fitting techniques (Lefever \cite{lefever}). Our sample is composed primarily of apparently slow rotators with low $\Omega R \sin i$ values not exceeding $\sim$65 km s$^{-1}$ (these figures are upper limits in the case of the $\beta$ Cephei variables, as the lines are often significantly broadened by pulsations). Our projected rotational velocities have been derived from profile fitting, but Fourier transform methods can also be used and have the advantage of disentangling different line-broadening mechanisms (Sim\'{o}n-D\'{\i}az \& Herrero \cite{simon_diaz}). Lefever (\cite{lefever}) used these techniques for 15 stars in our sample and found systematically lower  $\Omega R \sin i$ values. Although the differences are small and typically amount to only a few km s$^{-1}$, as expected considering that dwarfs suffer little macroturbulent broadening (Ebbets \cite{ebbets}), this bolsters our conclusion regarding the low rotation rate of our targets. Most importantly, asteroseismic and line-profile variation studies indicate that 10 stars have true rotational velocities well below 100 km s$^{-1}$ or, in some cases, as low as a few km s$^{-1}$ (\neri, \vcen \ and \tsco). As detailed in Table~\ref{tab_parameters}, this quantity is derived either from: a) identifying the recurrence timescale of the changes affecting the UV wind lines, with the rotational period in the 4 objects displaying phase-locked variations, or b) identification of the components of a rotationally split mode or theoretical modelling of the line-profile variations in the 6 pulsating variables.

\begin{table*}
\centering
\caption{Adopted atmospheric parameters and (projected) rotational velocities. The MK spectral types are based on our $T_{\rm eff}$ estimates and the calibrations for dwarfs of Crowther (\cite{crowther}). The calibration for supergiants was used for \icma. The 1-$\sigma$ uncertainty on $T_{\rm eff}$ is 1000 K for all stars. The quoted uncertainty on $\Omega R\sin i$ is the line-to-line scatter.}
\label{tab_parameters}
\begin{tabular}{rclcccccccc} \hline\hline
\multicolumn{1}{c}{HD}  & Usual  & Spectral type  & Remarks$^a$ & $T_{\rm eff}$     & $\log g$      & $\xi$ & $\Omega R\sin i$$^b$ & Ref. & $\Omega R$$^c$ & Ref. \\
\multicolumn{1}{c}{number}  & name    &     &     & [K]        & [cm s$^{-2}$] & [km s$^{-1}$] & [km s$^{-1}$] & & [km s$^{-1}$] & \\\hline
      886 & \gpeg       & B1.5--B2 IV   & SB1, $\beta$ Cephei & 22\,500  & 3.75$\pm$0.15  & 1$^{+2}_{-1}$ & 10$\pm$1 & 1 &                &    \\
     3360 & \zcas       & B2 IV         & SPB                 & 22\,000  & 3.70$\pm$0.15  & 1$\pm$1       & 19$\pm$1 & 2 & 55$\pm$28 (UV) &  6 \\
    16582 & \dcet       & B1.5--B2 IV   & $\beta$ Cephei      & 23\,000  & 3.80$\pm$0.15  & 1$^{+3}_{-1}$ & 14$\pm$1 & 1 & 14 or 28 (S)   &  7 \\
    29248 & \neri       & B1.5--B2 IV   & $\beta$ Cephei      & 23\,500  & 3.75$\pm$0.15  & 10$\pm$4      & 36$\pm$3 & 1 & 6 (S)          &  8 \\
    30836 & \piori      & B2 III        & SB1                 & 21\,500  & 3.35$\pm$0.15  &  8$\pm$4      & 43$\pm$3 & 3 &                &    \\
    35468 & \gori       & B2 IV--III    &                     & 22\,000  & 3.50$\pm$0.20  & 13$\pm$5      & 51$\pm$4 & 3 &                &    \\
    36591 &             & B1 V          &                     & 27\,000  & 4.00$\pm$0.20  & 3$\pm$2       & 16$\pm$2 & 3 &                &    \\
    44743 & \bcma       & B1.5 IV--III  & $\beta$ Cephei      & 24\,000  & 3.50$\pm$0.15  & 14$\pm$3      & 23$\pm$2 & 1 & 31$\pm$5 (LPV) &  9 \\
    46328 & \xcma       & B0.5--B1 IV   & $\beta$ Cephei      & 27\,500  & 3.75$\pm$0.15  & 6$\pm$2       & 10$\pm$2 & 1 &                &    \\
    50707 & \15cma      & B1 IV--III    & $\beta$ Cephei      & 26\,000  & 3.60$\pm$0.15  & 7$\pm$3       & 45$\pm$3 & 3 &                &    \\
    51309 & \icma       & B2.5 II       &                     & 17\,500  & 2.75$\pm$0.15  & 15$\pm$5      & 32$\pm$3 & 3 &                &    \\
    52089 & \ecma       & B1.5--B2 III  &                     & 23\,000  & 3.30$\pm$0.15  & 16$\pm$4      & 28$\pm$2 & 3 &                &    \\
    85953 &             & B2 IV         & SPB                 & 21\,000  & 3.80$\pm$0.15  & 1$\pm$1       & 29$\pm$2 & 2 &                &    \\
 111\,123 & \bcru       & B0.5--B1 IV   & SB1, $\beta$ Cephei & 27\,500  & 3.65$\pm$0.20  & 11$\pm$3      & 48$\pm$3 & 3 &                &    \\
 129\,929 & \vcen       & B1.5 V        & $\beta$ Cephei      & 24\,500  & 3.95$\pm$0.20  & 6$\pm$3       & 15$\pm$2 & 1 & 2 (S)          & 10 \\
 149\,438 & \tsco       & O9.5 V        &                     & 31\,500  & 4.05$\pm$0.15  & 2$\pm$2       &  8$\pm$2 & 4 & 6 (UV)         & 11 \\
 157\,056 & \toph       & B1.5 V        & SB1, $\beta$ Cephei & 25\,000  & 4.10$\pm$0.15  & 4$_{-3}^{+2}$ & 31$\pm$2 & 5 & 29 (S)         &  5 \\
 163\,472 & \voph       & B1.5--B2 V    & $\beta$ Cephei      & 23\,000  & 4.00$\pm$0.20  & 1$^{+4}_{-1}$ & 62$\pm$3 & 1 & 56 (UV)        & 12 \\
 205\,021 & \bcep       & B1 IV         & SB1, $\beta$ Cephei & 26\,000  & 3.70$\pm$0.15  & 6$\pm$3       & 29$\pm$2 & 1 & 26 (UV)        & 13 \\
 214\,993 & \twelvelac  & B1.5 IV       & $\beta$ Cephei      & 24\,500  & 3.65$\pm$0.15  & 10$\pm$4      & 42$\pm$4 & 1 & 45 (LPV)       & 14 \\
\hline
\end{tabular}
\begin{flushleft}
Key to references: [1] M06; [2] Briquet \& Morel (\cite{briquet_morel}); [3] this paper; [4] Hubrig et al., in prep.; [5] Briquet \etal (\cite{briquet}); [6] Neiner \etal (\cite{neiner03a}); [7] Aerts \etal (\cite{aerts06}); [8] Pamyatnykh \etal (\cite{pamyatnykh}); [9] Mazumdar \etal (\cite{mazumdar}); [10] Dupret \etal (\cite{dupret}); [11] Donati \etal (\cite{donati06b}); [12] Neiner \etal (\cite{neiner03b}); [13] Henrichs \etal (\cite{henrichs}); [14] Aerts (\cite{aerts96}).\\
$^a$ We only classify as $\beta$ Cephei stars the confirmed candidates in the catalogue of Stankov \& Handler (\cite{stankov_handler}).\\
$^b$ Note that the spectral lines are significantly broadened by pulsations in the $\beta$ Cephei and SPB stars (see M06). This explains why the quoted $\Omega R\sin i$ values, which reflect the total amount of line broadening arising from pulsations {\em and} rotation, are occasionally larger than the true rotational velocities, $\Omega R$.\\
$^c$ UV: calculated from the stellar radius derived from evolutionary tracks and assuming that the rotational period can be identified with the recurrence timescale of the changes affecting the UV wind lines, S: from asteroseismic studies, LPV: from modelling of the line-profile variations arising from pulsations.
\end{flushleft}
\end{table*}

\section{Reliability of the abundance results}\label{sect_sensitivity}
The CNO abundances for the complete sample are provided in Table~\ref{tab_abundances_cno}, along with the NLTE literature values for boron. Most B-depleted stars only have an upper limit to their boron abundance and we only considered in that case the most stringent estimate in the literature. Table~\ref{tab_abundances} presents the He, Mg, Al, Si, S and Fe abundances for the stars analyzed in the present study. Before discussing the chemical properties of our targets, we first assess the robustness of our results against the adopted atmospheric parameters. 

\begin{table*}
\centering
\caption{Mean NLTE B and CNO abundances (on the scale in which $\log \epsilon$[H]=12). The number of used lines is given in brackets. We define [N/C] and [N/O] as $\log$[$\epsilon$(N)/$\epsilon$(C)] and $\log$[$\epsilon$(N)/$\epsilon$(O)], respectively. The last column indicates the chemical group (Groups I, II or III) to which the star belongs to (see Sect.~\ref{sect_results} and Fig.\ref{fig_boron}).}
\hspace*{-0.7cm}
\label{tab_abundances_cno}
\begin{tabular}{rcccccccccc} \hline\hline
\multicolumn{1}{c}{HD number}  & Usual name & $\log \epsilon$(B) & Ref. & $\log \epsilon$(C) & $\log \epsilon$(N) & $\log \epsilon$(O) & ${\rm [N/C]}$ & ${\rm [N/O]}$ & Ref. & Group\\
\hline
      886 & \gpeg       &  2.23$\pm$0.20 & 1 & 8.20$\pm$0.05 (8)  & 7.58$\pm$0.11 (23) & 8.43$\pm$0.28 (21) & --0.62$\pm$0.12 & --0.85$\pm$0.30 & 5 & I\\
     3360 & \zcas       &  $<$1.21       & 1 & 8.16$\pm$0.08 (6)  & 7.97$\pm$0.13 (20) & 8.38$\pm$0.30 (18) & --0.19$\pm$0.15 & --0.41$\pm$0.32 & 6 & III\\ 
    16582 & \dcet       &  1.16$\pm$0.15 & 2 & 8.09$\pm$0.08 (6)  & 8.05$\pm$0.11 (26) & 8.45$\pm$0.26 (19) & --0.04$\pm$0.14 & --0.40$\pm$0.29 & 5 & III\\
    29248 & \neri       &  2.45$\pm$0.40 & 1 & 8.24$\pm$0.12 (10) & 7.87$\pm$0.09 (18) & 8.51$\pm$0.24 (19) & --0.37$\pm$0.15 & --0.64$\pm$0.26 & 5 & I\\
    30836 & \piori      &  $<$1.0        & 3 & 8.19$\pm$0.09 (5)  & 7.54$\pm$0.15 (13) & 8.40$\pm$0.29 (24) & --0.65$\pm$0.18 & --0.86$\pm$0.33 & 7 & II\\
    35468 & \gori       &  $<$1.01       & 1 & 8.11$\pm$0.09 (6)  & 7.90$\pm$0.16 (14) & 8.16$\pm$0.27 (12) & --0.20$\pm$0.19 & --0.26$\pm$0.32 & 7 & III\\
    36591 &             &  $<$1.27       & 4 & 8.19$\pm$0.05 (8)  & 7.66$\pm$0.12 (22) & 8.62$\pm$0.12 (27) & --0.53$\pm$0.13 & --0.95$\pm$0.17 & 7 & II\\
    44743 & \bcma       &  2.76$\pm$0.20 & 1 & 8.16$\pm$0.11 (9)  & 7.59$\pm$0.14 (26) & 8.62$\pm$0.18 (30) & --0.57$\pm$0.18 & --1.03$\pm$0.23 & 5 & I\\
    46328 & \xcma       &  $<$1.36       & 1 & 8.18$\pm$0.12 (9)  & 8.00$\pm$0.17 (34) & 8.59$\pm$0.17 (34) & --0.18$\pm$0.21 & --0.59$\pm$0.24 & 5 & III\\
    50707 & \15cma      &  $<$1.5        & 2 & 8.18$\pm$0.10 (5)  & 8.03$\pm$0.15 (23) & 8.64$\pm$0.18 (33) & --0.15$\pm$0.19 & --0.61$\pm$0.24 & 7 & III\\
    51309 & \icma       &  $<$0.84       & 1 & 7.89$\pm$0.19 (4)  & 7.86$\pm$0.34 (15) & 8.25$\pm$0.44 (11) & --0.03$\pm$0.39 & --0.39$\pm$0.56 & 7 & III\\
    52089 & \ecma       &  $<$1.62       & 1 & 8.09$\pm$0.12 (7)  & 7.93$\pm$0.14 (26) & 8.44$\pm$0.18 (29) & --0.16$\pm$0.19 & --0.51$\pm$0.23 & 7 & III\\
    85953 &             &                &   & 8.16$\pm$0.14 (9)  & 7.66$\pm$0.20 (11) & 8.41$\pm$0.32 (16) & --0.50$\pm$0.24 & --0.75$\pm$0.37 & 6 & \\ 
 111\,123 & \bcru       &  2.48$\pm$0.20 & 1 & 8.04$\pm$0.10 (7)  & 7.61$\pm$0.17 (19) & 8.59$\pm$0.16 (28) & --0.43$\pm$0.20 & --0.98$\pm$0.24 & 7 & I\\
 129\,929 & \vcen       &                &   & 8.34$\pm$0.13 (8)  & 7.73$\pm$0.10 (21) & 8.49$\pm$0.24 (25) & --0.61$\pm$0.17 & --0.76$\pm$0.26 & 5 & \\
 149\,438 & \tsco       &                &   & 8.19$\pm$0.14 (15) & 8.15$\pm$0.20 (35) & 8.62$\pm$0.20 (42) & --0.04$\pm$0.25 & --0.47$\pm$0.29 & 8 & \\  
 157\,056 & \toph       &                &   & 8.32$\pm$0.09 (7)  & 7.78$\pm$0.10 (23) & 8.58$\pm$0.26 (27) & --0.54$\pm$0.14 & --0.80$\pm$0.28 & 9 & \\
 163\,472 & \voph       &                &   & 8.21$\pm$0.07 (4)  & 7.99$\pm$0.17 (10) & 8.39$\pm$0.30 (14) & --0.22$\pm$0.19 & --0.40$\pm$0.35 & 5 & \\
 205\,021 & \bcep       &  $<$0.90       & 2 & 8.02$\pm$0.10 (11) & 7.91$\pm$0.13 (19) & 8.47$\pm$0.14 (30) & --0.11$\pm$0.17 & --0.56$\pm$0.19 & 5 & III\\
 214\,993 & \twelvelac  &  2.1$\pm$0.2   & 3 & 8.22$\pm$0.12 (6)  & 7.64$\pm$0.18 (17) & 8.42$\pm$0.23 (22) & --0.58$\pm$0.22 & --0.78$\pm$0.29 & 5 & I\\
\hline
\end{tabular}
\begin{flushleft}
Key to references: [1] Proffitt \& Quigley (\cite{proffitt_quigley}); [2] Venn \etal (\cite{venn02}); [3] Mendel \etal (\cite{mendel}); [4] Brooks \etal (\cite{brooks}); [5] M06; [6] Briquet \& Morel (\cite{briquet_morel}); [7] this paper; [8] Hubrig et al., in prep.; [9] Briquet \etal (\cite{briquet}).\\
\end{flushleft}
\end{table*}

\begin{table*}
\centering
\caption{Mean NLTE He, Mg, Al, Si, S and Fe abundances (on the scale in which $\log \epsilon$[H]=12) for the stars analyzed in the present study. The data for the remaining objects in the sample can be found in the references in Table~\ref{tab_abundances_cno}. The number of used lines is given in brackets. For comparison purposes, the first line gives the typical values found for B dwarfs in the solar neighbourhood (M06; Daflon \& Cunha \cite{daflon_cunha}; Gummersbach \etal \cite{gummersbach}; Kilian-Montenbruck \etal \cite{kilian_montenbruck}).}
\label{tab_abundances}
\begin{tabular}{rccccccc} \hline\hline
HD number     & Usual name & He/H & $\log \epsilon$(Mg) & $\log \epsilon$(Al) & $\log \epsilon$(Si) &  $\log \epsilon$(S) &  $\log \epsilon$(Fe)\\
\hline
         & B dwarfs & $\sim$0.085 & $\sim$7.4   & $\sim$6.1   & $\sim$7.2   & $\sim$7.2   & $\sim$7.3\\
  30836  & \piori   & 0.074$\pm$0.034 (9)  & 7.33$\pm$0.34 (1)  & 6.06$\pm$0.14 (4)  & 7.20$\pm$0.33 (9)  & 7.18$\pm$0.21 (5)  & 7.11$\pm$0.18 (19)\\
  35468  & \gori    & 0.088$\pm$0.024 (6)  & 7.16$\pm$0.26 (1)  & 5.98$\pm$0.08 (3)  & 7.00$\pm$0.19 (7)  & 7.06$\pm$0.21 (3)  & 7.23$\pm$0.14 (11)\\ 
  36591  &          & 0.059$\pm$0.016 (9)  & 7.33$\pm$0.20 (1)  & 6.21$\pm$0.15 (3)  & 7.25$\pm$0.29 (10) & 7.06$\pm$0.21 (1)  & 7.32$\pm$0.19 (31)\\ 
  50707  & \15cma   & 0.100$\pm$0.024 (7)  & 7.34$\pm$0.20 (1)  & 6.16$\pm$0.17 (3)  & 7.31$\pm$0.30 (8)  & 7.17$\pm$0.21 (1)  & 7.23$\pm$0.19 (16)\\
  51309  & \icma    & 0.047$\pm$0.020 (7)  & 7.30$\pm$0.39 (1)  & 5.92$\pm$0.21 (3)  & 7.18$\pm$0.31 (9)  & 6.99$\pm$0.19 (8)  & 7.18$\pm$0.30 (10)\\ 
  52089  & \ecma    & 0.078$\pm$0.018 (5)  & 7.39$\pm$0.23 (1)  & 6.08$\pm$0.13 (3)  & 7.23$\pm$0.24 (10) & 7.16$\pm$0.25 (4)  & 7.16$\pm$0.15 (22)\\
111\,123 & \bcru    & 0.067$\pm$0.020 (9)  & 7.30$\pm$0.20  (1) & 6.18$\pm$0.20  (4) & 7.25$\pm$0.23  (9) & 7.14$\pm$0.24  (1) & 7.23$\pm$0.24 (17)\\
\hline
\end{tabular}
\end{table*}

Of particular concern is the uncertainty surrounding the determination of the microturbulent velocity in the giants. As already largely discussed in the literature (e.g. Daflon \etal \cite{daflon}; Kilian \cite{kilian92}; Trundle \etal \cite{trundle04}), larger microturbulence values are usually obtained for the more evolved objects ($\xi$ and $\log g$ are tightly anticorrelated in our dataset). Furthermore, discrepant values are obtained depending on the chemical species chosen (e.g. the \ion{O}{ii} features yielding systematically larger values than the \ion{N}{ii} lines). Fortunately, the prime indicators for an N enrichment, namely the ratios of the C, N and O abundances ([N/C] and [N/O]), are largely insensitive to this problem. In the case of \ecma, for instance, these ratios differ by a negligible amount of at most 0.04 dex when adopting the microturbulence estimated from the oxygen ($\xi$=16 km s$^{-1}$) or nitrogen lines ($\xi$=12 km s$^{-1}$). The microturbulent velocity is also poorly constrained in \icma \ because of the lack of strong ionic features in this temperature range, but once again this does not affect our conclusions.

\begin{figure*}
\centering
\includegraphics[width=15cm]{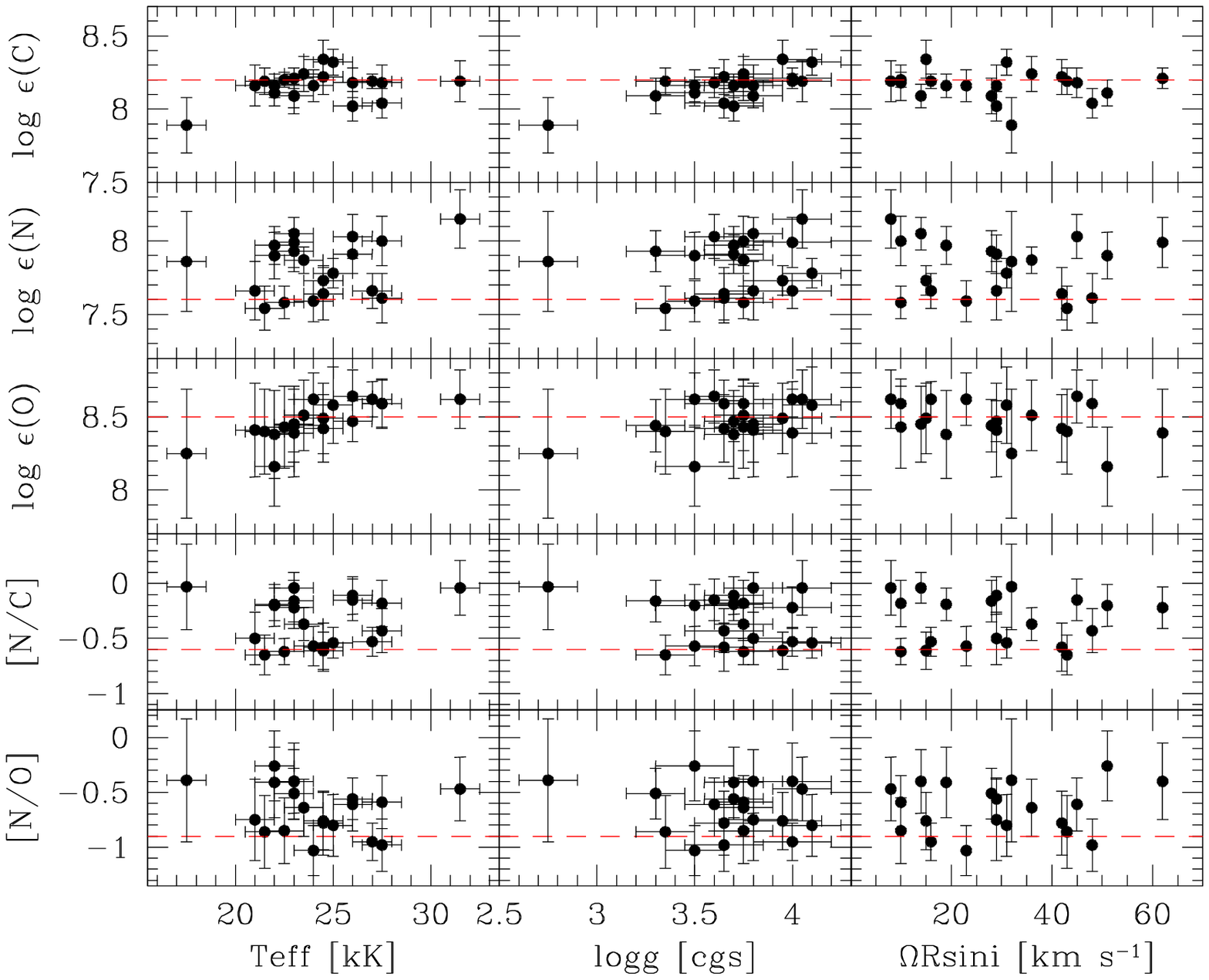}
\caption{Variations of the CNO abundances and their ratios, as a function of the effective temperature, surface gravity and projected rotational velocities. The dashed lines indicate the typical values found for nearby B dwarfs (M06; Daflon \& Cunha \cite{daflon_cunha}; Gummersbach \etal \cite{gummersbach}; Kilian-Montenbruck \etal \cite{kilian_montenbruck}). Similar [N/C] and [N/O] ratios are found for the Sun (Grevesse \& Sauval \cite{grevesse_sauval}; Asplund \etal \cite{asplund}), and can thus be considered as the baseline values in the absence of rotational mixing.}
\label{fig_trends_CNO}
\end{figure*}

\begin{figure}
\centering
\includegraphics[width=6.5cm]{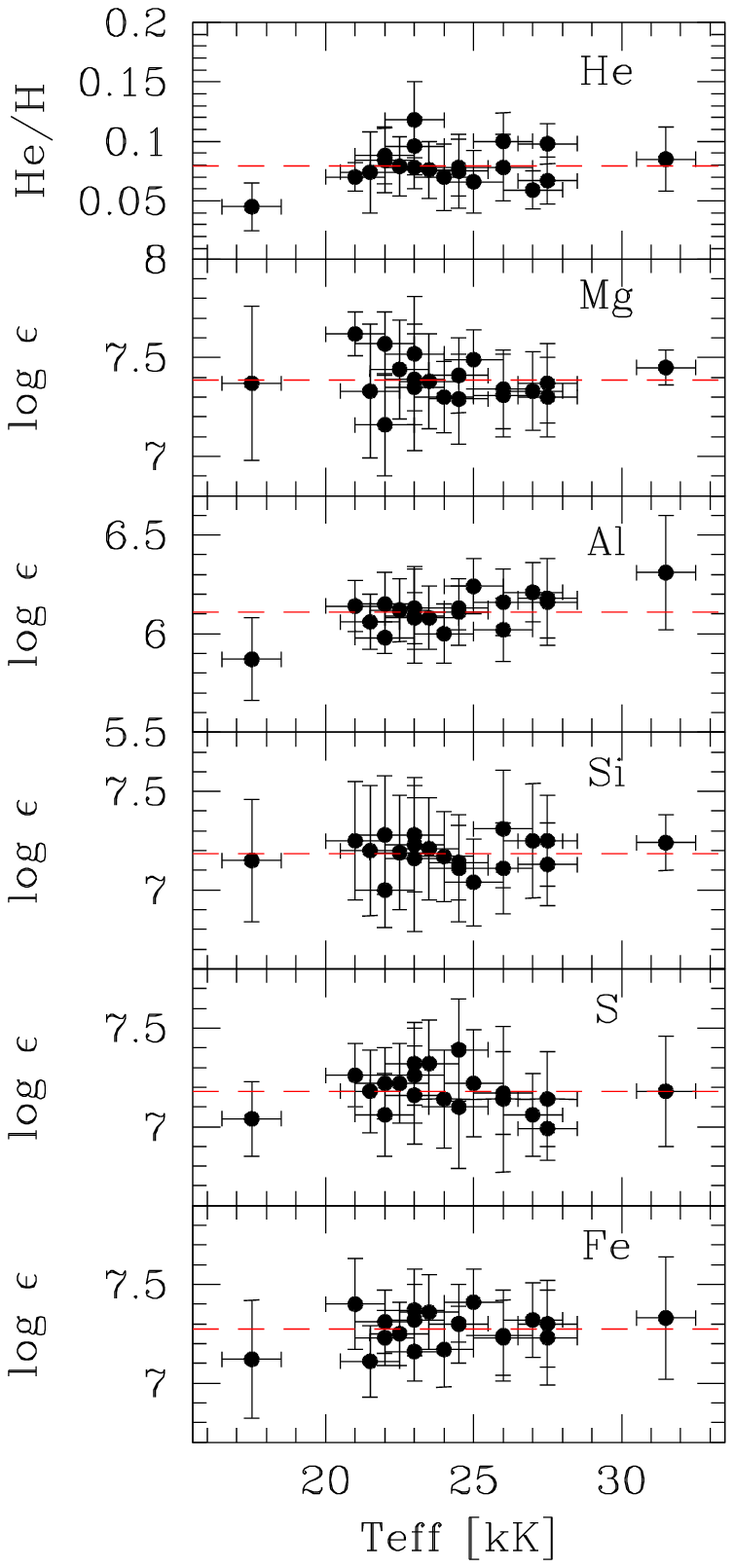}
\caption{Variations of the He, Mg, Al, Si, S and Fe abundances, as a function of the effective temperature. The mean values are indicated by dashed lines.}
\label{fig_trends_He_Fe}
\end{figure}

Figure~\ref{fig_trends_CNO} displays the variations of the CNO abundances and their ratios, as a function of the atmospheric parameters and projected rotational velocity. The oxygen abundance becomes increasingly difficult to measure with confidence as one progresses towards later spectral types because of its strong sensitivity to errors on $T_{\rm eff}$ and the intrinsic weakness of the \ion{O}{ii} features, but there is evidence for lower abundance values as $T_{\rm eff}$ decreases. The lack of similar trends, either for carbon and nitrogen, or for helium and the other dominant metals (except Al; see Fig.\ref{fig_trends_He_Fe}) suggests that this problem is most likely intrinsic to oxygen (perhaps related to the NLTE corrections) and not attributable to an incorrect temperature scale. Because of this temperature-dependence affecting oxygen and the fact that carbon is a more sensitive indicator of core-processed material brought up to the surface (the predicted depletion factors are higher than for oxygen), we will primarily focus on the [N/C] ratio in the following (furthermore, the uncertainties affecting the C abundances are a factor $\sim$2 lower; Table~\ref{tab_abundances_cno}). 

There is also a discernible correlation between the C abundances and $\log g$ (Fig.\ref{fig_trends_CNO}). If we use the surface gravity as a proxy for the evolutionary status, this would interestingly point to a progressive depletion of carbon at advanced, stellar ages as a consequence of CNO-cycle burning. However, such a C depletion in the most evolved objects should be systematically counterbalanced by an N excess, which is not observed. In addition, the C depletion factors (based on the N excesses discussed below) are expected to be of the order of the uncertainties ($\sim$0.1 dex) and, as such, barely detectable (there is indeed no anticorrelation between the C and N abundances). This, along with the fact that the [N/C] and [N/O] ratios show no trends with $\log g$, argues against the existence of evolutionary effects in our data (we will return to this issue in Sect.~\ref{sect_population}). 

We note that uncertain, but generally low abundances are obtained for He and most metals in the case of the B2.5 II star \icma \ (Figures \ref{fig_trends_CNO} and \ref{fig_trends_He_Fe}). These low values might indicate a general problem with our analysis of this star (neglect of wind effects being a possibility) and turn out to contribute for a large part to the correlations discussed above.

Due to our inability, as others, to satisfactorily model some strong \ion{C}{ii} lines (most notably \ion{C}{ii} $\lambda$4267), we based our abundance determination for carbon on a number of weak lines (see M06 for the complete line list). Nieva \& Przybilla (\cite{nieva}) recently developed a carbon model atom that allowed them to overcome this problem and to obtain consistent results for all lines they considered. A straight comparison with their results for the two stars in common shows that our values are lower by 0.14 and 0.11 dex for \object{HD 36591} and \tsco, respectively. However, the net effect of the model atom can be better appraised by recomputing the abundances using their adopted atmospheric parameters and measured EWs. For the lines in common, we now find mean values lower by 0.06 and higher by 0.02 dex for \object{HD 36591} and \tsco, respectively. Our ability to closely reproduce the results of Nieva \& Przybilla (\cite{nieva}) validates the use of our model atom. The systematically higher abundances they found cannot be explained by the slight differences in the atmospheric parameters, but rather by the fact that their observed line profiles are apparently stronger (their EWs are on average $\sim$15\% higher than ours), perhaps as a result of the different continuum rectification.

The boron abundances collected from the literature have been derived from the spectral synthesis of the \ion{B}{iii} $\lambda$2065.8 resonance line. This feature is strongly blended with metallic lines (e.g. \ion{Mn}{iii} $\lambda$2065.9) and is usually weak in the objects suffering significant depletion. The bulk of the boron data is derived from Proffitt \& Quigley (\cite{proffitt_quigley}) who used archival {\it International Ultraviolet Explorer (IUE)} high-dispersion SWP spectra. They mention that excellent fits were achieved for $T_{\rm eff}$ $\lesssim$ 28\,000 K, which is the case for all stars with boron data in our sample. Contrary to \ion{B}{ii} $\lambda$1362, for instance, the NLTE corrections for \ion{B}{iii} $\lambda$2066 are modest in the $T_{\rm eff}$ range of interest (typically --0.15 dex) and should not constitute a major source of error. Although it is difficult to assess the reliability of these abundance determinations, subsequent studies based on higher quality data obtained with the STIS spectrograph onboard the {\it Hubble Space Telescope (HST)} support these estimates (Brooks \etal \cite{brooks}; Venn \etal \cite{venn02}; Mendel \etal \cite{mendel}). Namely, the same results within the uncertainties were found in cases of detection (e.g. \twelvelac), while more stringent upper limits were set for stars with very weak lines (e.g. \15cma \ and \bcep). These more recent values were preferred and are those quoted in Table~\ref{tab_abundances_cno}.

\section{Results}\label{sect_results}
\begin{figure*}
\centering
\includegraphics[width=18.5cm]{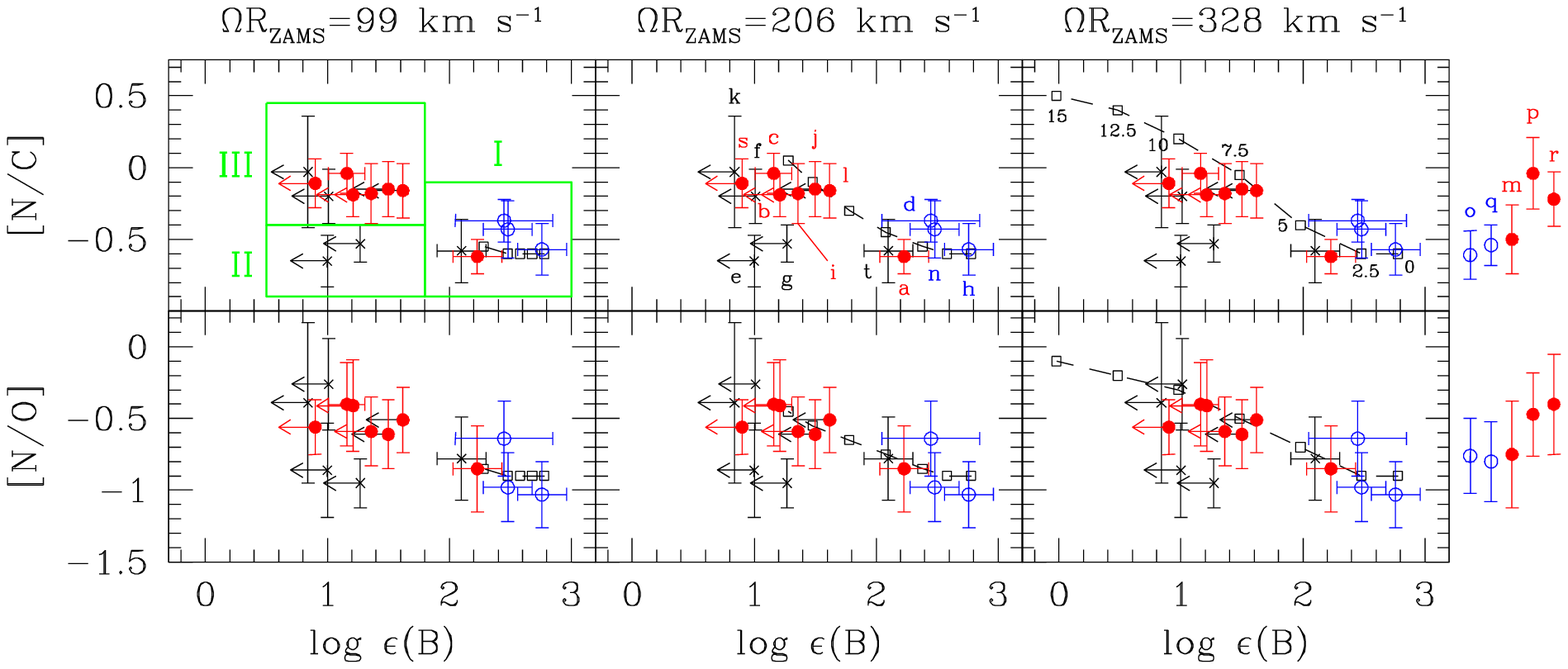}
\caption{[N/C] and [N/O] ratios as a function of the NLTE B abundances: ({\em a}) \gpeg, ({\em b}) \zcas, ({\em c}) \dcet, ({\em d}) \neri, ({\em e}) \piori, ({\em f}) \gori, ({\em g}) \object{HD 36591}, ({\em h}) \bcma, ({\em i}) \xcma, ({\em j}) \15cma, ({\em k}) \icma, ({\em l}) \ecma, ({\em m}) \object{HD 85953}, ({\em n}) \bcru, ({\em o}) \vcen, ({\em p}) \tsco, ({\em q}) \toph, ({\em r}) \voph, ({\em s}) \bcep \ and ({\em t}) \twelvelac. {\em Filled circles}: stars with a detected magnetic field; {\em open circles}: stars with null detections; {\em crosses}: stars without magnetic data (in the online version of this journal, red and blue circles refer to stars with and without a detected magnetic field, respectively). Five stars lacking boron data, but which have been the subject of magnetic searches are shown off scale to the right-hand side of this figure. The arrows denote upper limits. The abundance data are compared with the theoretical predictions of Heger \& Langer (\cite{heger_langer}) for a 12 M$_{\sun}$ star and three different values of the equatorial rotational velocity on the ZAMS: $\Omega R_{\rm ZAMS}$=99 ({\em left-hand panels}), 206 ({\em middle panels}) and 328 km s$^{-1}$ ({\em right-hand panels}). The locus in each panel ({\em dashed line and open squares}) defines an age sequence with the time elapsed from the ZAMS increasing leftwards: from $t$=0 to 15 Myrs (0 to 12.5 Myrs for $\Omega R_{\rm ZAMS}$=99 km s$^{-1}$) in steps of 2.5 Myrs (see upper, right-hand panel). The initial [N/C], [N/O], and boron abundances at $t$=0 have been taken as --0.6, --0.9 (representative of the pristine solar value: Grevesse \& Sauval \cite{grevesse_sauval}; Asplund \etal \cite{asplund}) and $\log \epsilon$(B)=2.78$\pm$0.04 dex (meteoritic value; Zhai \& Shaw \cite{zhai_shaw}), respectively. The boxes in the upper, left-hand panel delineate the three classes of stars (Groups I, II and III) with different chemical properties (see text).}
\label{fig_boron}
\end{figure*}

Figure~\ref{fig_boron} shows the behaviour of the [N/C] and [N/O] ratios, as a function of the boron content. The targets are plotted using different symbols according to the detection, or lack thereof, of a magnetic field. Five stars lacking boron data, but which have been searched for magnetic fields are plotted off scale to the right-hand side of this figure. Three groups of stars with distinct chemical properties, and which may define an evolutionary sequence, can be identified (see boxes in the upper, left-hand panel of Fig.\ref{fig_boron}) and are discussed in turn below (see also Table~\ref{tab_abundances_cno} for a classification of the targets in the different groups).

{\bf Group I --} First, 5 stars exhibit [N/C] and [N/O] ratios close to --0.6 and --0.9, respectively. These values are nearly identical to those found for the Sun, either using classical 1-D model atmospheres (Grevesse \& Sauval \cite{grevesse_sauval}) or time-dependent, 3-D hydrodynamical models incorporating updated atomic data and NLTE corrections (Asplund \etal \cite{asplund}). They can thus be considered as the baseline values in the absence of deep mixing. This is supported by the lack, or relatively mild, boron depletion compared with the meteoritic value ($\log \epsilon$[B]=2.78 dex; Zhai \& Shaw \cite{zhai_shaw}). These stars have not yet undergone substantial deep mixing and boron is only beginning to be transported to deeper layers of the stellar envelope where it is destroyed by proton capture. No clear case for an N-enriched star with a solar B abundance can be found (the N overabundance of \neri \ and \bcru \ is only marginal), in accordance with the theoretical expectations.

{\bf Group II --} Two stars (\piori \ and \object{HD 36591}) present an interesting behaviour: they are very boron depleted (by a factor of at least 40), yet have solar nitrogen abundances. Shallow mixing has dramatically depleted the photospheric boron, but deep mixing is either absent or, at least, has yet to bring detectable amounts of CN-cycle burning products to the surface. As can be seen in Fig.\ref{fig_boron}, the chemical properties of these two stars are not well-reproduced by theoretical models (Fliegner \etal \cite{fliegner}; Heger \& Langer \cite{heger_langer}). Specifically, the surface B depletion and N enrichment do not seem to proceed gradually and in parallel, as predicted, but that the latter is delayed and may only begin after boron has been depleted by $\sim$1.5 dex. 

{\bf Group III --} Finally, another population (8 stars) exhibits very low boron abundances (indeed mostly upper limits) coupled with an N enrichment reaching a factor about 3--4 (the separation between the two subsamples of N-normal and N-rich stars is clearer in terms of the [N/C] ratio, which is the most robust diagnostic for an N excess). The abundance patterns of this subgroup are well-reproduced by the theoretical models of Heger \& Langer (\cite{heger_langer}), but only for initial velocities exceeding $\sim$200 km s$^{-1}$ (see Fig.\ref{fig_boron}).\footnote{The Geneva models do not give predictions for the boron abundances, but they lead to even lower N enrichments (Meynet \& Maeder \cite{meynet_maeder03}).} This is well above the true rotation rates of \zcas, \dcet \ and \bcep: $\Omega R$ $\lesssim$ 60 km s$^{-1}$. The other N-rich stars have $\Omega R \sin i$ $\lesssim$ 50 km s$^{-1}$ (Table~\ref{tab_parameters}). We will argue below that these stars only suffered a moderate amount of angular momentum loss along the main sequence and that they were likely to be already slow rotators on the ZAMS.

\section{Discussion}\label{sect_conclusion}
\subsection{A population of slowly-rotating, N-rich Galactic B dwarfs}\label{sect_population}
Our NLTE abundance analysis confirms the results of previous studies (e.g. Kilian \cite{kilian92}; Gies \& Lambert \cite{gies_lambert}) and supports the existence of a population of B stars which, despite having low rotation rates, displays the observational signature of deep mixing during the early phases of their evolution. Of the 36 (non-supergiant) B stars observed by Gies \& Lambert (\cite{gies_lambert}), about one third exhibit a [N/C] ratio above about --0.4 dex. Such a high incidence of N enriched stars in a primarily magnitude limited (with an additional constraint on the projected rotational velocity) sample suggests that such objects make up a significant fraction of all B dwarfs in the solar neighbourhood. A sizeable, similar population has also been shown to exist in the Magellanic Clouds (Hunter \etal \cite{hunter07a},b; Trundle \etal \cite{trundle07}). 

Mass transfer processes in close interacting binaries may also dramatically alter the CNO surface abundances of the two components (e.g. Vanbeveren \cite{vanbeveren}) and may be responsible for the existence of the so-called OBN stars whose unusual chemical properties are believed to be connected to binarity (e.g. Levato \etal \cite{levato}). However, one expects in that case a much higher N overabundance and C depletion than observed in our sample (Sch\"onberner \etal \cite{schonberner}). A good example of such a star with a more extreme CNO abundance pattern is the blue straggler \object{$\theta$ Car} (Hubrig et al., in prep.). On the other hand, \bcep \  is the only N-rich star known to be member of a binary system. The orbit is very wide (${\cal P}_{\rm orb}$$\sim$90 yr; Pigulski \& Boratyn \cite{pigulski_boratyn}) and the existence of a past episode of mass transfer is debatable. In the case of \piori, the strong boron depletion but roughly normal nitrogen abundance have been claimed to be the unambiguous signature of rotational mixing (Fliegner \etal \cite{fliegner}). Furthermore, three other single-lined binary systems (\gpeg, \ \bcru \ and \toph) have normal abundances. Summarizing, mass exchange processes do not appear as a viable explanation for all the N overabundances observed. 

\begin{figure*}
\centering
\includegraphics[width=16cm]{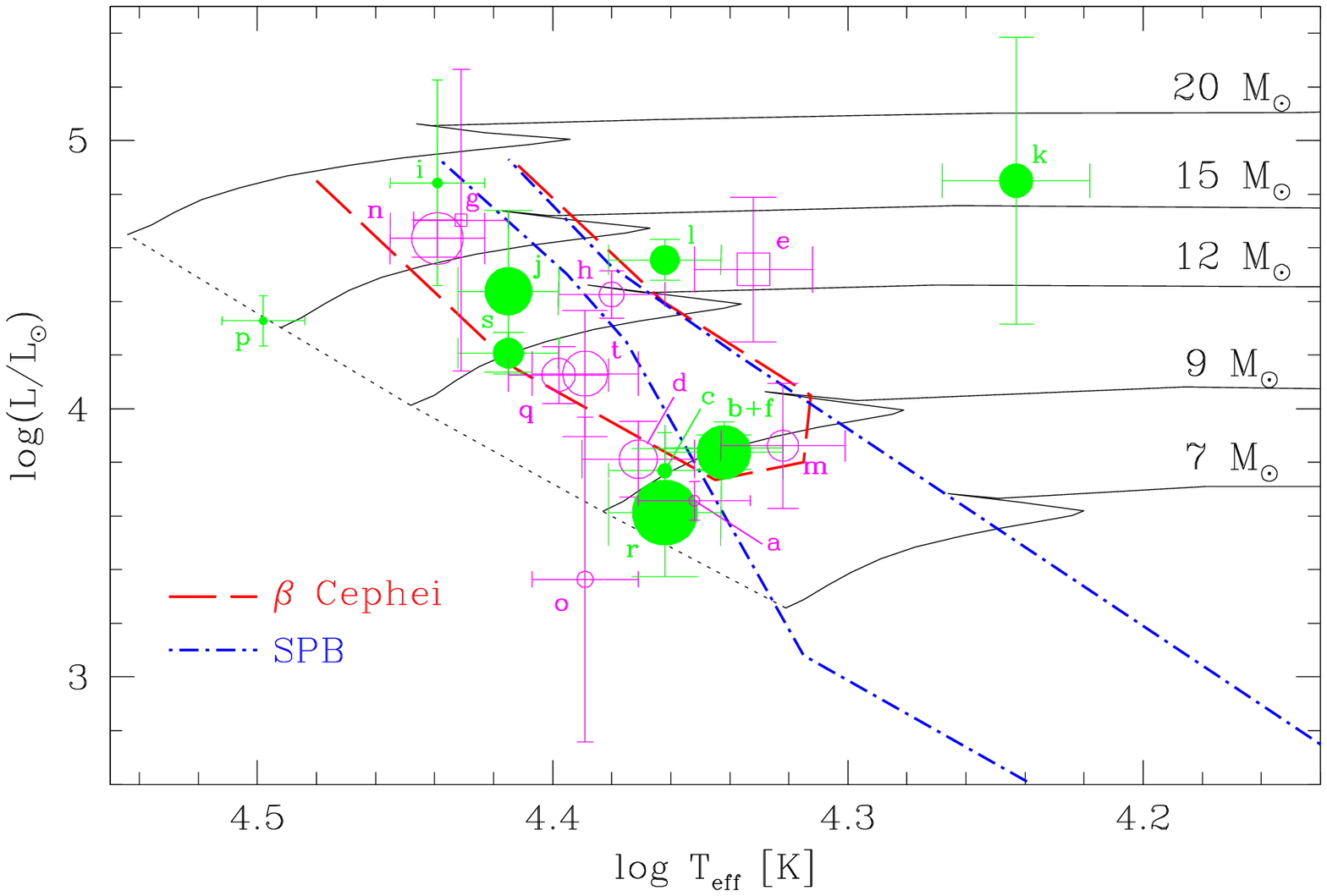}
\caption{Position of the programme stars in the Hertzsprung-Russell diagram: ({\em a}) \gpeg, ({\em b}) \zcas, ({\em c}) \dcet, ({\em d}) \neri, ({\em e}) \piori, ({\em f}) \gori, ({\em g}) \object{HD 36591}, ({\em h}) \bcma, ({\em i}) \xcma, ({\em j}) \15cma, ({\em k}) \icma, ({\em l}) \ecma, ({\em m}) \object{HD 85953}, ({\em n}) \bcru, ({\em o}) \vcen, ({\em p}) \tsco, ({\em q}) \toph, ({\em r}) \voph, ({\em s}) \bcep \ and ({\em t}) \twelvelac. {\em Open circles}: Group I (B- and N-normal stars), {\em open squares}: Group II (N-normal, but B-depleted stars), {\em filled circles}: Group III (B-depleted and N-rich stars). See Fig.\ref{fig_boron} for a definition of the three subgroups. In the online version of this journal, green and magenta symbols refer to stars with and without an N excess, respectively. The stars lacking boron data (see right-hand side of Fig.\ref{fig_boron}) are included in this figure assuming that the 3 N-normal and 2 N-rich stars belong to Groups I and III, respectively. The size of the symbols is linearly proportional to the $\Omega R \sin i$ values (see Table~\ref{tab_parameters}). The luminosities were computed using {\em Hipparcos} parallaxes and the bolometric corrections of Flower (\cite{flower}). The extinction in the $V$ band, $A_V$, was derived from the theoretical ($B-V$) colour indices of Bessell \etal (\cite{bessell}). The evolutionary tracks for solar metallicity and without rotation are taken from Schaller \etal (\cite{schaller}). The initial masses are indicated on the right-hand side of this figure and the ZAMS is shown as a dotted line. The theoretical instability strips for $\beta$ Cephei and SPB stars are shown as long dashed and dotted-dashed lines, respectively (Miglio \etal \cite{miglio}). The models have been computed assuming a metallicity $Z$=0.01 (value typical of the $\beta$ Cephei stars in our sample; see M06), updated OP opacities, the solar mixture of Asplund \etal (\cite{asplund}) and the mean Ne abundance of Cunha \etal (\cite{cunha}).}
\label{fig_hr}
\end{figure*}

With the caveat that not all the N-rich stars in our sample have been the subject of intensive monitoring campaigns, one can note that at least two stars in this subsample (\gori \ and \ecma) are classified as poor or rejected $\beta$ Cephei candidates in the recent catalogue of Stankov \& Handler (\cite{stankov_handler}). As can be seen in Fig.\ref{fig_hr}, two N-rich stars (\icma \ and \tsco) also fall well outside the instability domains for pressure and gravity modes.\footnote{A blue loop evolution can be safely ruled out for the evolved star \icma \ according to evolutionary models (both with and without rotation included). The observed abundance anomalies are indeed much less severe than expected if the star had gone through the first dredge-up (e.g. Venn \cite{venn99}).} On the other hand, some well-known pulsators are N-normal (e.g. \gpeg, \bcma). This argues against a purely pulsational origin for the nitrogen enrichment.

On the other hand, except perhaps in the case of \icma, the stellar winds are too weak to remove the outermost stellar layers and expose boron-poor material. As an illustration, mass-loss rates $\dot{M}$$\sim$10$^{-7}$ M$_{\sun}$ yr$^{-1}$ are required (e.g. Venn \etal \cite{venn02}), whereas two of the most evolved stars in our sample fall short by at least one order of magnitude: $\dot{M}$$\sim$6 and 9 $\times$ 10$^{-9}$ M$_{\sun}$ yr$^{-1}$ for \bcma \ and \ecma, respectively (Cassinelli \etal \cite{cassinelli}; Drew \etal \cite{drew}). The same conclusion holds for the main-sequence star \tsco, with $\dot{M}$$\sim$2 $\times$ 10$^{-8}$ M$_{\sun}$ yr$^{-1}$ (Donati \etal \cite{donati06b}).

It can be argued that some stars in our sample were rapid rotators on the ZAMS, hence experiencing strong rotational mixing and surface N enrichment, but that substantial loss of angular momentum ensued, spinning down the stars to the observed levels. This hypothesis is not supported by evolutionary models, which only predict a slow decline of the rotation rate during core-hydrogen burning (e.g. Meynet \& Maeder \cite{meynet_maeder03}), as confirmed by observations of early-B dwarfs in young open clusters (e.g. Huang \& Gies \cite{huang_gies}; Wolff \etal \cite{wolff}). However, the case of the magnetic stars is more controversial. For instance, a magnetically-confined wind could constitute an effective way of dissipating angular momentum (as in the case of the O star \object{HD 191612}; Donati \etal \cite{donati06a}). The influence of a large-scale magnetic field with a dipole configuration on a radiatively-driven outflow can be parametrized by a dimensionless confinement parameter,
\begin{equation}
\eta=B^2_{\rm pol}R^2_{\star}/\dot{M}v_{\infty}
\end{equation}
where $B_{\rm pol}$ is the polar field strength, $R_{\star}$ is the stellar radius, $\dot{M}$ is the mass-loss rate and $v_{\infty}$ is the terminal wind velocity (ud-Doula \& Owocki \cite{ud_doula}). The emergent flow is channeled along closed magnetic loops  and deflected towards the magnetic equatorial plane for $\eta$ $\gtrsim$ 1. Assuming an oblique rotator model for \zcas, \voph \ and \bcep \ leads to a polar field of about 300 G (Neiner \etal \cite{neiner03a},b; Henrichs \etal \cite{henrichs}). The field geometry is more complex in \tsco, but the average field emerging at the surface typically reaches similar values (Donati \etal \cite{donati06b}). If we adopt, as representative values, $B_{\rm pol}$=300 G, $R_{\star}$=5 R$_{\odot}$, $\dot{M}$=10$^{-8}$ M$_{\sun}$ yr$^{-1}$ (see Sect.~\ref{sect_population}) and $v_{\infty}$=1500 km s$^{-1}$, we obtain $\eta$$\sim$30. It is thus likely that the stellar outflow would be magnetically confined within the Alfv\'en radius, $R_A$, in our magnetic targets. Following Donati \etal (\cite{donati01}), the magnetic braking timescale, $\tau$, can be roughly estimated through the following expression
\begin{equation}
\tau=k\frac{M_{\star}}{\dot{m}}\left(\frac{R_{\star}}{R_A}\right)^2
\end{equation}
where $k$ is the fractional gyration radius, $M_{\star}$ is the stellar mass and $\dot{m}$ is the effective mass-loss rate (i.e. the amount of wind material not trapped within the magnetosphere). Characteristic values for $\dot{m}$ and $R_A$ have been obtained in the case of \bcep \ and \tsco \ by Donati \etal (\cite{donati01}, \cite{donati06b}) using the magnetically-confined wind-shock model of Babel \& Montmerle (\cite{babel_montmerle}). This led to magnetic braking timescales that are much longer than the estimated stellar ages (110 Myrs and 5 Gyrs, respectively). There is considerable uncertainty surrounding $\dot{m}$ and the loss of angular momentum through magnetic phenomena is not fully understood (for instance, some chemically peculiar Bp stars undergo unexpectedly strong, sometimes abrupt, increases in their rotational period: Mikul\'{a}\v{s}ek \etal \cite{mikulasek}; Trigilio \etal \cite{trigilio}), but it appears unlikely that the slow rotation rate of the N-rich stars results from magnetic braking. Another argument against this interpretation is the fact that some non-magnetic stars in our sample are also intrinsically slow rotators (Table~\ref{tab_parameters}). This suggests that another mechanism could also be at work, a possibility being that this property is inherited from the pre-main sequence phase (e.g. St\c{e}pie\'n \cite{stepien}; Wolff \etal \cite{wolff}).

\subsection{A link with magnetic fields?}\label{sect_fields}
Rotational mixing effects are expected to be more conspicuous at lower metallicities and for increasing stellar mass, age and rotation rate (e.g. Meynet \& Maeder \cite{meynet_maeder00}). There are no trends between the abundance data and $\Omega R\sin i$ (Fig.\ref{fig_trends_CNO}) or $\Omega R$, but this is somewhat expected considering the limited range in rotation rate sampled by our sample (and the noise introduced by the unknown inclination angles in the case of the projected rotational velocities). More telling is the lack of evidence for the stars with an N excess to be on average older and/or more massive than the stars with normal N abundances (Fig.\ref{fig_hr}). As can also be seen in Fig.\ref{fig_boron}, all the N-rich stars reach very similar (``saturation''?) levels of N enhancement ($\sim$0.5 dex) despite having different physical properties. The inadequacy of the theoretical models to reproduce our observations is well-illustrated by the unexpected detection of a factor of 3 nitrogen excess in \tsco \ and \voph. First, these stars are very slow rotators (especially \tsco \ with $\Omega R$$\sim$6 km s$^{-1}$; Donati \etal \cite{donati06b}) and rotational mixing should be particularly inefficient in that case. This problem could be lessened if one assumes a much steeper internal rotation law than currently assumed, leading to an increased transport of the chemical elements through shear instabilities. A decline of the rotational velocity by a factor 3--5 from the core to the surface has been inferred from asteroseismological studies in  \neri \ (Pamyatnykh \etal \cite{pamyatnykh}) and \vcen \ (Dupret \etal \cite{dupret}). Although this information is unfortunately lacking for the N-rich stars, this appears in accordance with the predictions of evolutionary models (Meynet \& Maeder \cite{meynet_maeder00}). Second, these stars are also probably very young (see Fig.\ref{fig_hr}) and should not have had time to build up a significant N overabundance at the surface. Taken together, this suggests that an additional physical process to rotational mixing might play a key role in the appearance of N-enriched material at the stellar surface. 

In this respect, it is remarkable that all observational data collected to date point to a higher incidence of a magnetic field in stars with peculiar abundances: all observations of the N-rich stars led to a magnetic field detection, whereas all stars with solar abundances remained undetected.\footnote{We do not discuss here the case of the N-rich, apparently slowly-rotating $\beta$ Cephei star \object{HD 180642} (Morel \& Aerts \cite{morel_aerts}), as the magnetic field detection is not completely secure and needs to be confirmed: although several Zeeman features are seen in Stokes {\em V} spectra obtained with FORS1/VLT, none of the two spectropolarimetric measurements formally exceeds the 3$\sigma$ detection threshold (Hubrig et al., in prep.).} The only two exceptions to this rule are \gpeg \ and \object{HD 85953} (Fig.\ref{fig_boron}). The $\beta$ Cephei star \gpeg \ has been claimed to possess a very weak magnetic field with a longitudinal component varying from --10 to 30 G (Butkovskaya \& Plachinda \cite{butkovskaya}). On the other hand, a magnetic field has been detected in the Slowly Pulsating B (SPB) star \object{HD 85953} at the 3-$\sigma$ level in 3 out of 5 measurements spread over 3.5 yrs with FORS1/VLT (Hubrig \etal \cite{hubrig06}; Hubrig et al., in prep.). The latter example illustrates the paucity of the magnetic fields measurements in B-type stars and that caution should be exercised when using these data. The field detections reported in Table~\ref{tab_magnetic_fields} are for the most part at the limit of the instrumental capabilities (especially with the older generation of spectropolarimeters such as MuSiCoS) and are based on snapshot observations with a poor temporal sampling. Only in the cases of \zcas, \tsco, \voph \ and \bcep \ have sufficient time-resolved measurements been collected to provide detailed information on the field strength and global morphology (Neiner \etal \cite{neiner03a},b; Donati \etal \cite{donati06b}; Henrichs \etal \cite{henrichs}). Considering that the null detections are also often based on a handful of measurements and might be refuted when more intensive and precise observations are undertaken, the separation between magnetic and non-magnetic stars should only be regarded as arbitrary at this stage. It is also conceivable that the N enrichment took place when the star was very young and going through a magnetic phase (similarly to \object{$\theta^1$ Ori C}), and that the field strongly decayed afterwards (e.g. Hubrig \etal \cite{hubrig07}) to a point where it is no longer detectable with current instrumental facilities. In this context, it is interesting to note that an N excess is already present in the young magnetic stars \tsco \ and \voph \ (see Fig.\ref{fig_hr}). Although highly speculative, this might explain why a slowly-rotating, N-rich dwarf such as \object{1 Cas} ([N/C]=--0.07$\pm$0.20 dex; Gies \& Lambert \cite{gies_lambert}) remained undetected during the intensive MuSiCoS observations of Schnerr (\cite{schnerr07}). A correlation between the amount of CN-cycle processed material at the surface and the field strength would strongly support a connection between deep mixing and magnetic phenomena. There is no trend in our data between [N/C] and the rms longitudinal field $\overline{\left< B_l \right>}$ (Table~\ref{tab_magnetic_fields}), but it should be kept in mind that the rotational period is poorly sampled in most cases and that the magnetic observations are restricted to longitudinal field measurements, which are aspect dependent, and to fields larger than $\sim$40--50 Gauss. Despite these limitations, however, forthcoming magnetic observations of OB stars conducted using the new generation of spectropolarimetric instruments (such as FORS1/VLT, ESPaDOnS/CFHT or Narval/TBL) will clearly prove essential to establish the existence of a link between a nitrogen enrichment and magnetic fields. Such a relationship could in principle help to select the most promising targets for magnetic field surveys of massive stars (indeed, an N excess was already one of the selection criteria used by Schnerr \cite{schnerr07}).

Strong, ordered kG-strength magnetic fields have been detected in two slowly-rotating O-type stars, namely in the O6.5f?pe--O8fp star \object{HD 191612} (Donati \etal \cite{donati06a}) and in the young, peculiar O7 star \object{$\theta^1$ Ori C} (Wade \etal \cite{wade}). To our knowledge, however, no quantitative CNO abundance analysis has been performed for either of these two objects. Only in the former has a nitrogen enhancement been suggested based on the observed line strengths (Walborn \etal \cite{walborn}). Unfortunately, a similar lack of CNO abundance data also applies to classical, magnetic Bp stars lying in a similar temperature domain as our programme stars. Two strongly magnetic, He-rich dwarfs (\object{HD 66522} and \object{HD 96446}) have been analyzed under the assumption of LTE by Zboril \& North (\cite{zboril}). The CNO abundances do not clearly differ from those of the non-magnetic stars in their sample, but no firm conclusion can be drawn from such a limited number of objects. Systematic studies have been conducted for Ap stars, both for boron (e.g. Leckrone \cite{leckrone}) and CNO (e.g. Roby \& Lambert \cite{roby}). However, these stars are much cooler than our targets and diffusion effects have long been known to play a key role in the dramatic surface abundance anomalies observed in this temperature range.

Nevertheless, radiatively-driven microscopic diffusion has also been proposed (Bourge \etal \cite{bourge07}) as a possible explanation for the appearance of core-processed material at the surface of slowly-rotating $\beta$ Cephei variables (we recall that 5 N-rich targets are confirmed members of this class; see Table~\ref{tab_parameters}). Support for the occurrence of diffusion comes from the fact that it may also account for the accumulation of iron in the driving zone and the excitation of the pulsation modes through the $\kappa$ mechanism (e.g. Bourge \etal \cite{bourge06}). Magnetic fields were not taken into account in these models, but they are expected to amplify the transport of the ionized species along the field lines and to lead to a large-scale, inhomogeneous distribution of some elements at the stellar surface. However, the Mg, Al, Si, S and Fe abundances do not reveal any clear deviation from a scaled solar pattern (see Fig.\ref{fig_trends_He_Fe}) and do not provide evidence for the existence  of diffusion effects capable of significantly altering the surface abundances. This is supported by a comparison of the abundance properties of the two subsamples of N-normal and N-rich stars, which reveals no significant differences for chemical elements other than nitrogen (Table~\ref{tab_comparison}). Two stars apparently have a low helium content (\object{HD 36591} and \icma), but this is at odds with the expected/observed position of the He weak stars in the $T_{\rm eff}$-$\log g$ plane (Babel \cite{babel}; Bychkov \etal \cite{bychkov}) and may be regarded with some suspicion. Slightly subsolar He abundances are in general derived within our sample (Fig.\ref{fig_trends_He_Fe}). We suspect that the adopted microturbulent velocities derived from the analysis of the \ion{O}{ii} features are not adequate for properly modelling the \ion{He}{i} lines, and that lower values should ideally be employed instead (as also concluded for \ion{N}{ii}; Sect.~\ref{sect_sensitivity}).

\begin{table}
\centering
\caption{Comparison between the chemical properties of the two subsamples of N-normal and N-rich stars.}
\label{tab_comparison}
\begin{tabular}{lcc} \hline\hline
        & N-normal sample & N-rich sample\\
        & (10 stars) & (10 stars)\\
\hline
$$He/H              & 0.071$\pm$0.007 & 0.087$\pm$0.020\\
$\log \epsilon$(C)  & 8.21$\pm$0.09   & 8.11$\pm$0.10\\ 
$\log \epsilon$(N)  & 7.67$\pm$0.11   & 7.98$\pm$0.09\\ 
$\log \epsilon$(O)  & 8.51$\pm$0.09   & 8.44$\pm$0.16\\ 
$\log \epsilon$(Mg) & 7.39$\pm$0.11   & 7.38$\pm$0.12\\ 
$\log \epsilon$(Al) & 6.13$\pm$0.08   & 6.10$\pm$0.12\\ 
$\log \epsilon$(Si) & 7.18$\pm$0.07   & 7.19$\pm$0.10\\ 
$\log \epsilon$(S)  & 7.20$\pm$0.11   & 7.15$\pm$0.11\\ 
$\log \epsilon$(Fe) & 7.28$\pm$0.10   & 7.26$\pm$0.08\\ 
\hline
\end{tabular}
\end{table}

Recent codes attempting to model rotational mixing in the presence of magnetic fields (Maeder \& Meynet \cite{maeder_meynet}; Heger \etal \cite{heger}) also face difficulties for accounting for the data presented in this paper. For instance, they predict a noticeable helium enrichment, which is not observed (the only exception is \voph, but the slight He excess is more likely to be due to magnetic phenomena; see Neiner \etal \cite{neiner03b} and M06). Furthermore, these models assume a magnetic field created via dynamo action in the stellar envelope (Spruit \cite{spruit}), whereas at least three N-rich stars in our sample (\zcas, \voph \ and  \bcep) have been claimed to possess a global dipole field most likely of fossil origin (Neiner \etal \cite{neiner03a},b; Henrichs \etal \cite{henrichs}). A remnant field from the star formation phase is also more likely in \tsco, although the field topology is probably much more complex (Donati \etal \cite{donati06b}). In all cases, the periods found in the magnetic data are compatible with those modulating the UV wind line profiles, thus supporting an oblique rotator model. Clearly much progress remains to be done before the production of magnetic fields in massive stars and the mixing processes taking place in their interiors are satisfactorily understood (see, e.g. Zahn \etal \cite{zahn}).

\begin{acknowledgements}
T. M. acknowledges financial support from the Research Council of Leuven University through grant GOA/2003/04. We are very grateful to the referee, P. Dufton, for useful comments and for providing us with the FEROS spectra of \icma \ and \ecma. We are also indebted to C. Aerts and M. Desmet for acquiring the CORALIE spectra of \piori, \15cma, \bcru \ and \tsco. We would like to thank K. Butler for making the NLTE line-formation codes DETAIL/SURFACE available to us, as well as A. Miglio and J. Montalb{\'a}n for useful discussions about the diffusion of helium. Valuable suggestions from C. Aerts were also very much appreciated. We are grateful to L. Decin and M. F. Nieva for their help on the construction of the model grid at high temperatures. The archival ELODIE data have been processed within the PLEINPOT environment. This research made use of NASA's Astrophysics Data System Bibliographic Services, the SIMBAD database operated at CDS, Strasbourg (France). 
\end{acknowledgements}


\end{document}